\documentclass[a4paper,11pt]{article}
\usepackage{pos}

\usepackage[all,2cell]{xy} \UseAllTwocells 							
\usepackage{color}
\definecolor{cmyk}{cmyk}{.0,.4,.9,.5}								    
\usepackage{slashed}
\usepackage{MnSymbol}
\usepackage{orcidlink}
   
\newcommand{\B}{\mathcal{B}}

\newcommand{\F}{\mathcal{F}}
\renewcommand{\H}{\mathcal{H}} 	
\newcommand{\K}{\mathcal{K}} 

\newcommand{\M}{\mathcal{M}}
 
\renewcommand{\O}{\mathcal{O}}	
 
\newcommand{\CC}{{\mathbb{C}}}

\newcommand{\HH}{{\mathbb{H}}}

\newcommand{\MM}{{\mathbb{M}}}

\newcommand{\RR}{{\mathbb{R}}} 

\newcommand{\WW}{{\mathbb{W}}}

\newcommand{\As}{{\mathscr{A}}}

\newcommand{\Hs}{{\mathscr{H}}}
 
\newcommand{\Os}{{\mathscr{O}}}

\DeclareFontFamily{U}{rsfs}{\skewchar\font127}
\DeclareFontShape{U}{rsfs}{m}{n}{%
   <5> <6> rsfs5
   <7> rsfs7
   <8> <9> <10> <10.95> <12> <14.4> <17.28> <20.74> <24.88> rsfs10
}{}
\DeclareSymbolFont{rsfs}{U}{rsfs}{m}{n} 
\DeclareSymbolFontAlphabet{\scr}{rsfs} 

\DeclareMathOperator{\WWy}{\WW y}

\DeclareMathOperator{\id}{Id}

\DeclareMathOperator{\re}{Re}
\DeclareMathOperator{\im}{Im}

\newcommand{\cj}[1]{\overline{#1}}								
\newcommand{\ip}[2]{\langle #1\mid #2\rangle}			
\renewcommand{\iff}{\Leftrightarrow}							
\newcommand{\imp}{\Rightarrow}										

\renewcommand{\logo}{\relax}                     

\title{Quantum Born Geometries from Modular Quantization}

\author*[a]{Paolo Bertozzini}
\author[b]{Roberto Conti}

\affiliation[a]{\orcidlink{0000-0002-1387-9283} Independent Researcher, 
Bangkok, Thailand}

\affiliation[b]{\orcidlink{0000-0003-3128-6884} Dipartimento di Scienze di Base e Applicate per l'Ingegneria, University,\\
Sapienza Universit\`a di Roma, Via A.~Scarpa 16, I-00161 Roma, Italy}

\emailAdd{paolo.th@gmail.com}
\emailAdd{roberto.conti@sbai.uniroma1.it}

\date{16 November 2026}

\abstract{ 
\begin{center}16 November 2026\end{center}
A structural link between quantization and {\it ``thermal-doubled'' non-commutative Born geometry} (with its {\it split-quaternionic structure}) is uncovered via {\it Tomita-Takesaki first-quantized modular theory} for complexified and polarized ortho-symplectic spaces. 

\medskip 

Implications for the {\it modular algebraic quantum gravity} program ({\href{https://arxiv.org/abs/1007.4094v1}{arXiv:1007.4094v1}}) hint at further deep modifications of the notions of space and geometry ({\it thermal quantum space-time hypothesis}) waiting to be further explored.

\medskip 

Speculative suggestions in the direction of A.Sakharov and L.Boyle-K.Finn-N.Turok {\it CPT-mirror cosmologies} will be mentioned, among several others. 

\medskip

The present work is based on the manuscript \cite{BC26} that has been shared, with a selected number of researchers, since end of April 2026. 

\bigskip 


\medskip 

MCS-2020: {\it
81S10, 			
53D18, 			
81T75, 			
83C45,			
81T05, 			
46L51, 			
46L87,			
53D50. 			
}
}

\FullConference{Proceedings of the Corfu Summer Institute 2026 
"Workshop on Noncommutative and Generalised Geometries and their Physical Applications" 
(CORFU2026) \\ 
13 September - 21 September 2026\\
Corfu, Greece\\}

\begin{document}

\maketitle 

\section{Strategy for Modular Algebraic Quantum Theory}

Contrary to the current prevailing view, that space-time is an ``emergent'' feature absent at the quantum gravity level, revisiting old and forgotten M.Born-P.Jordan's suggestions \cite{Ca21}, we would like to repropose, in the light of first-quantized modular theory that:  
\begin{itemize}
\item[$\rightsquigarrow$]\footnote{The $\rightsquigarrow$ symbol is used in this paper to conveniently denote certain statements that we deem to be essential messages.} {\it
quantization of phase-space, even in ordinary quantum theory, already entails a quantization of configuration-space;  
hence to the question 
``Is there geometry at the fundamental quantum level?''  
our answer will be 
``Yes, modular non-commutative geometry!''. }
\end{itemize}
Following basic ideas in \cite{BCL10} \cite{Ra17a,Ra17b} \cite{B23,BC26}, in the {\it modular algebraic quantum gravity} program:
\begin{itemize}
\item[]
we aim to revert the implication \ 
$[\text{classical geometry}]\imp [\text{local equilibrium states}]$, 
\\ 
and obtain \emph{geometry from states} \ 
$[\text{quantum geometry}] \Leftarrow [\text{local equilibrium states}]$. 
\end{itemize}

We stress that, contrary to the usual interpretation of modular-time as a ``classical macroscopic time'' following the {\it thermal-time hypothesis} \cite{CoRo94}, what we propose (see \cite{B23} for more motivation in this direction) is a fully covariant {\it thermal-spacetime hypothesis} recovering a spacetime quantum geometry, from the modular theory induced by a ``state'' on covariant partial observables. 

\medskip 

In the classical case ($1^{\text{st}}$ line in the diagram below) the equivalence between geometry and equilibrium states is beautifully described, via Katz-Gallavotti KMS property and Weinstein modular theory (see \cite[section 3]{Wei97} \cite[section 2.3]{Cic09} \cite[remark 12]{DrWa21} \cite[section 10.4]{Ne26}). 
{\footnotesize
\begin{equation*}
\xymatrix{
\boxed{
\text{
\begin{minipage}{1.57cm}
\begin{center}
classical \\ 
geometry
\end{center}
\end{minipage}
}
} 
\ar@{~>}[d]|{\text{modular quantization}}
\ar@<+0.5ex>[rrr]^{
\quad \quad \text{
\begin{minipage}{2.6cm} 
\begin{center}
free geodesic motion 
\\[-0.2cm]
on phase-space 
\end{center}
\end{minipage}
}
\quad \quad 
}
& & & 
\ar@<+0.5ex>[lll]^{
\quad \quad 
\text{
\begin{minipage}{2.7cm} 
\begin{center}
polarization 
\\[-0.2cm]
localization/transport 
\end{center}
\end{minipage}
}
\quad \quad 
} 
\boxed{
\text{
\begin{minipage}{1.37cm} 
\begin{center}
classical 
\\
dynamics
\end{center}
\end{minipage}
}
} 
\ar@<+0.5ex>[rrr]^{
\text{
\begin{minipage}{2.0cm} 
\begin{center}
\text{Katz-Gallavotti}
\\[-0.2cm]
\text{KMS-property}
\end{center}
\end{minipage}
}
\quad \quad 
}
& & & 
\ar@<+0.5ex>[lll]^{
\text{
\begin{minipage}{2.1cm} 
\begin{center}
\text{Weinstein}
\\[-0.2cm]
\text{modular theory}
\end{center}
\end{minipage}
}
\quad \quad 
} 
\boxed{\text{
\begin{minipage}{2.4cm} 
\begin{center}
classical (local) 
\\
equilibrium states
\end{center}
\end{minipage}
}
}
\ar@{~>}[d]|{\text{algebraic (Mauro Weyl) quantization}}
\\ 
\boxed{
\text{\begin{minipage}{1.57cm} 
\begin{center} 
quantum 
\\ 
geometry
\end{center}
\end{minipage}
}
}\  
& & & \ 
\boxed{
\text{\begin{minipage}{1.38cm} 
\begin{center}
quantum 
\\ 
dynamics
\end{center}
\end{minipage}
}
}
\ar@{~>}[lll]_{\text{
\begin{minipage}{2.6cm} 
\begin{center} 
\text{modular localization}
\\[-0.2cm]
\text{(descent)}
\end{center}
\end{minipage}
\ }
}|{\text{polarization}}^{
\text{
\begin{minipage}{2.3cm} 
\begin{center}
\text{modular transport}
\\[-0.2cm]
\text{(holonomy)}
\end{center}
\end{minipage}
}}   
\ar@<+0.5ex>[rrr]^{
\text{
\begin{minipage}{1.9cm} 
\begin{center}
\text{KMS-property}
\end{center}
\end{minipage}
}
\quad \quad 
}
& & & 
\ar@<+0.5ex>[lll]^{
\text{
\begin{minipage}{2.1cm} 
\begin{center}
\text{Tomita-Takesaki}
\\[-0.2cm]
\text{modular theory}
\end{center}
\end{minipage}
}
\quad \quad 
} 
\boxed{\text{
\begin{minipage}{2.4cm} 
\begin{center} 
quantum (local) 
\\
equilibrium states
\end{center}
\end{minipage}
}
}
}
\end{equation*}
} 

\noindent 
In the quantum case ($2^{\text{nd}}$ line in the above diagram), via Tomita-Takesaki theory, certain modularly localized Liouvillian spectral geometries (non-commutative geometries at the level of phase-space) have been described in \cite{BCL10}; but the impossibility to identify, from modular theory alone, a suitable {\it polarization}, prompted the ``reverse-engineering'' efforts discussed in \cite{BC26} and here: 
starting with a classical Riemannian space, at the first-quantization level, we will identify the geometric significance of the modular ingredients and obtain (upon a choice of polarization) quantum Born geometries. 
The insight will remain valid in the Lorentzian situation as well as non-second-quantized cases. 
 
\subsection[(Some) Basic Physical Principles in Modular Algebraic Quantum Theory]
{
(Some) Basic Physical Principles in Modular Algebraic Quantum Theory\footnote{Only the last 3 points below will be directly treated in this work.}
}
\begin{itemize}
\item[$\rightsquigarrow$] 
\emph{Modular localization}:
quantum subsystems are determined by states via the KMS-condition, 
\item[$\rightsquigarrow$] 
\emph{Jacobi-Einstein generalized equivalence principle}: 
all interactions are subsumed in a background geometry and dynamics is free subject only to the geometric constraints, 
\item[$\rightsquigarrow$] 
\emph{Geometric source of modular theory}:  
(first-quantized) modular theory originates in complexification of ortho-symplectic spaces and encodes free dynamics contraints on phase-spaces, 
\item[$\rightsquigarrow$] 
\emph{Modular CPT-symmetry}: 
quantum systems are in CPT-symmetric thermal-double pairs,  
\item[$\rightsquigarrow$] 
\emph{Modular Born geometry}: 
induced by a CPT symmetry $J$, 
a modular duality $L$, 
a Born reciprocity $\Phi$, 
a modular $\CC$-polarization $\Gamma$,  
a (choice of) modular Born $\RR$-polarization $\Lambda$. 
\end{itemize}

\section[Background on Tomita-Takesaki Modular Theory]{Background on Tomita-Takesaki Modular Theory \cite{Su06} \cite{Ta03}} 
\begin{itemize}
\item[$\rightsquigarrow$]
{\it the missing secret ingredient for ``quantization''}.
\end{itemize}
For a von Neumann algebra $\M=\M''\subset \B(\H)$ on a Hilbert space $\H$, 
for every vector state $\phi_\xi(T):=\ip{\xi}{T(\xi)}$, with $\xi\in \H$, that is 
{\it cyclic}: $\cj{(\M \xi)}=\H$, and {\it separating}: $A\xi=0 \imp A=0$, 
the $\RR$-subspace 
$\{T(\xi) \ | \ T=T^*, \ T\in\M \}\subset \H$ 
is standard (see property~\eqref{eq: standard} below) and we have that: 
\begin{itemize}
\item
its modular one-parameter unitary group 
$t \mapsto \Delta_\xi^{it}=\exp(it K_\xi)\in \B(\H)$, 
\item
its modular conjugation antilinear isometry $\H \xrightarrow{J_\xi} \H$, satisfy:
\begin{gather*}
\sigma^\phi_t(\M):=\Delta_\xi^{it}\M\Delta_\xi^{-it}= \M, \quad \forall t \in \RR,
\quad \quad 
J_\xi\M J_\xi = \M', 
\end{gather*}
where the \emph{commutant} $\M'$ of $\M$ is defined by: 
$\M':=\{A' \in \B(\H) \ | \ [A',A]_{-}=0, \forall A \in \M \}$.\footnote{ 
Notice that $\M'$ is a von Neumann algebra and $\M''=\M$.} 
\end{itemize}

Here below we will exclusively make use of modular theory in its first-quantized version. 

\subsection[Modular Theory - First Quantized Level]{
Modular Theory - First Quantized Level \cite{RvD77} \cite{Lo08,Lo21} \cite{Ne25} \cite{CLL26}
}

Any complex Hilbert space $\H_\CC$ with a \emph{standard real subspace} $V_\RR$: 
\begin{equation} \label{eq: standard}
\H_\CC=\cj{V_\RR\oplus i\cdot V_\RR}, \quad \quad V_\RR\cap i\cdot V_\RR=\{0_\H\},
\end{equation}
canonically determines the following data: 
\begin{itemize}
\item 
a complex unitary structure $T_V:\xi\mapsto i\cdot \xi$, for $\xi\in \H_\RR$, 
\item 
a densely defined Tomita conjugation $S_V:v\oplus (i\cdot w)\mapsto v\oplus (-i\cdot w), 
\quad x,y\in V_\RR$,  
\\ 
with polar Hilbert decomposition $\cj{S}_V=J_V\circ\Delta_V^{\frac{1}{2}}$, 
\item 
a unitary one-parameter group $t\mapsto \Delta_V^{it}=\exp(it K_V)$ on $\H_\CC$, leaving $V_\RR$ stable: 
\begin{equation*}
\Delta_V^{it}(V_\RR)=V_\RR, \quad \forall t\in\RR,
\end{equation*}  
\item
an anti-unitary operator $J_V$ mapping $V_\RR$ on its {\it symplectic complement}  
\begin{equation*}
V':=\{h\in \H_\CC \ | \ \im\ip{h}{v}_\H=0, v\in V_\RR \}.
\end{equation*} 
\item[$\rightsquigarrow$] \it 
The physical nature of the modular ortho-symplectic dynamics $t\mapsto \Delta_V^{it}$ and of the symmetry $J_V$ is still generally unknown! 
\end{itemize}

\subsection[Algebraic Quantum Field Theory, Free Scalar Fields, Reeh-Schlieder Theorem]{
Algebraic Quantum Field Theory, Free Scalar Fields, Reeh-Schlieder Theorem
\cite{Ar63,Ar64,Wa94,BGP07} 
}

\begin{center}
\boxed{
\begin{minipage}{3.5cm}
\begin{center}
globally hyperbolic 
Lorentzian submanifold
$(\O,g)$ 
\end{center}
\end{minipage}
}
$\imp$
\boxed{
\begin{minipage}{4.5cm}
\begin{center}
symplectic $\RR$-space $(\Gamma_\O,\omega)$
of local solutions of Klein-Gordon-equation $\square_g\phi=m\phi$
\end{center}
\end{minipage}
}
$\imp$ 
\boxed{
\begin{minipage}{3.5cm}
\begin{center}
complexified 
Weyl-algebra 
$\As(\O):=\WWy(\Gamma_\O,\omega)$ 
\end{center}
\end{minipage}
}
\end{center}
\begin{itemize}
\item 
the functor $\O\mapsto\As(\O)$ is an algebraic quantum field theory, 
\item 
the complexification $\Gamma_\O^\CC$ of $(\Gamma_\O,\omega)$ is a Kre\u\i n space, 
\item 
a choice of K\"ahler $\CC$-polarization (equivalently a quasi-free state on $\WWy(\Gamma_\O,\omega)$) selects a one-particle Hilbert space $\Hs^+$. 
\end{itemize}
By H.Reeh-S.Schlieder theorem \cite{RS61}:
whenever the open region $\Os\neq\varnothing$ has a non-trivial causal complement $\Os'\neq\varnothing$,
the Bosonic Fock space $\Hs:=\F_-(\Hs^+)$ of $\Hs^+$ is a standard Hilbert space for the von Neumann algebra $\As(\O)''$ generated by $\As(\O)$, with vacuum state vector $\xi_o\in\Hs$ that is cyclic for both $\As(\O)$ and its commutant $\As(\O)'$ and hence also separating for $\As(\O)''$.  
\begin{itemize}
\item[$\rightsquigarrow$] 
\emph{there is a mysterious modular dynamics on each $\As(\O)''$ that we unconventionally conjecture to be a ``Coriolis' effect'' of a non-commutative geometry of space-time. 
}
\end{itemize}

\section[Modular Quantization]{Modular Quantization \cite{BC26}} 

\begin{itemize}
\item[$\rightsquigarrow$]
{\it quantization of phase-space entails, via modular theory, also a quantization of configuration and momentum spaces}. 
\end{itemize}

\subsection[External Ortho-symplectic Complexifications]{
External Ortho-symplectic Complexifications \cite[section 3.2]{BC26}
}

By \emph{ortho-symplectic $\RR$-space}\footnote{
Notice that in the K\"ahler case $\omega_V(v,w)=g_V(i\cdot v,w)$, for all $v,w\in V$, and so $\|\omega_V\|=1$. 
}
$(V_\RR,g_V,\omega_V)$ we mean a $\RR$-vector space $V_\RR$ with an inner product $g_V$ and an anti-symmetric bilinear map $\omega_V$ such that $\|\omega_V\|< 1$. 

The \emph{external complexification} of $(V_\RR,g_V,\omega_V)$ is given by the Hilbert completion $\H_\CC$ of $V_\RR\oplus V_\RR$, with $\CC$-action given, for all $\alpha,\beta\in \RR$ and $v,w\in V$, by:  
\begin{equation*}
(\alpha+i\beta)\cdot (v\oplus w):= (\alpha v-\beta w)\oplus(\alpha w+\beta v), 
\end{equation*}
and the positive sesquilinear form given, for all $v_1\oplus w_1,v_2\oplus w_2\in V_\RR\oplus V_\RR$, by: 
\begin{multline*}
h(v_1\oplus w_1,v_2\oplus w_2):=
\big(+g_V(v_1,v_2) + g_V(w_1,w_2) -\omega_V(v_1,w_2) + \omega_V(w_1,v_2)\big)
\\
+i\cdot \big(+g_V(v_1,w_2) -g_V(w_1,v_2) +\omega_V(v_1,v_2) + \omega_V(w_1,w_2)\big).
\end{multline*} 
There is a categorical equivalence: 
$\xymatrix{\boxed{
\begin{minipage}{4.0cm}
\begin{center}
complexifications of \\ ortho-symplectic spaces 
\end{center}
\end{minipage}
}
\ar@{<:>}[r]
&
\boxed{
\begin{minipage}{4cm}
\begin{center}
standard $\RR$-subspaces of $\CC$-Hilbert spaces 
\end{center}
\end{minipage}
}}$ 
\begin{itemize}
\item[$\rightsquigarrow$] \it 
and hence any ortho-symplectic space is equipped with a modular flux and, inside its external complexification,  it is mapped by a modular conjugation $J_V$ onto its symplectic complement. 
\end{itemize}

\subsection[Gauge Modular Flow of a Riemannian Manifold]{
Gauge Modular Flow of a Riemannian Manifold \cite[section 3.4]{BC26} \cite{NaZh25,Na25} 
}

The previous insight can be applied to any generalized tangent space of a Riemannian manifold: 
\begin{gather*}
\xymatrix{
\boxed{\text{\begin{minipage}{3.5cm} \begin{center}
Riemannian manifold \\ $(X,g_X)$\end{center}\end{minipage}}} \ar@{:>}[rr]
& & 
\boxed{\text{\begin{minipage}{4.5cm} \begin{center} modular gauge groups 
$\sigma^{g_X,\gamma}_q, \ \gamma\in]0,1[, \ q\in X$ 
\end{center} \end{minipage}}}
}
\end{gather*}
a Riemannian classical geometry $(X,g_X)$, for $0<\gamma<1$, has canonically associated (first-quantized) gauge modular flows $\sigma^{g_X,\gamma}$ on the 
\emph{external complexification 
$\H_X:=(TX\oplus T^*X)\otimes_\RR\CC$ of its Hitchin-Gualtieri \cite{Hi10} \cite{Gu04} generalized tangent bundle $V_X:=TX\oplus T^*X$}: 
\begin{itemize}
\item 
the Hitchin-Gualtieri generalized tangent bundle $V_X:=TX\oplus T^*X$ is a K\"ahler ortho-symplectic bundle with symplectic structure $\omega_V(x_1\oplus y^*_1,x_2\oplus y^*_2):=y^*_2(x_1)-y^*_1(x_2)$ and inner product $g_V:=g_X\oplus g_{X^*}$, with $g_{X^*}(x_1^*,x_2^*):=g_X(x_1,x_2)$, for all $x_j,y_j\in TX$ and $x_j^*,y_j^*\in T^*X$, $j=1,2$, where we denoted by $x\mapsto x^*$, for $x\in TX$, the Riesz isomorphism between $TX$ and $T^*X$, 
\item 
for all $0<\gamma<1$, $V_X$ becomes an ortho-symplectic bundle with symplectic form $\gamma\cdot \omega_V$ and metric $g_V$ and hence a standard $\RR$-subbundle of the Hilbert bundle $\H_X$ with inner product $h^g+i\gamma h^\omega$ where $h^g$ and $h^\omega$ are the complexified sesquilinear forms induced by $g_V$ and $\omega_V$. 
\end{itemize}
Notice that in absence of the symplectic form, for $\gamma\to 0$, the modular flux $\sigma^{g_X,\gamma}$ is ``freezing'' to the trivial identity map; for $\gamma\to 1$, in the K\"ahlerian case, the modular flux is ``divergent''. 
\begin{itemize}
\item[$\rightsquigarrow$] \it 
what is the physical/geometrical meaning of the modular gauge groups and conjugations:
\begin{equation*}
t\mapsto\sigma^{g_X,\gamma}_{t}:=\Delta^{it}_{V_X,g_X,\gamma}, \quad \text{and} \quad J^{g_X,\gamma}_{V_X} \ ?
\end{equation*}
\end{itemize}

\subsection[$\RR$-polarization, Duality, Born Reciprocity]{
$\RR$-polarization, Duality, Born Reciprocity \cite{Bo38} \cite[section 3.1]{BC26}
}

When $M=T^*X$ we have a global \emph{$\RR$-polarization} of the phase-space $M$ and, given a metric $(X,{g_X})$, the Ehresmann Levi-Civita connection on $T^*X$ induces a canonical decomposition:  
\begin{equation*}
T_{m}M\simeq T_{\pi(m)}X\oplus T^*_{\pi(m)}X, \quad \forall m\in M; \quad R:x\oplus y^*\mapsto x\oplus(-y^*).
\end{equation*} 
The bundle $V:=TM$ decomposes as direct sum of Riesz-dual Lagrangian spaces $TM\simeq TX\oplus T^*X$ with 
$\RR$-polarization $R$.
The symplectic form $\omega_V$ induces (modulo sign) a canonical \emph{Born symmetry} $TM\xrightarrow{B}TM$ (given by $(x,y^*)\mapsto(-y,x^*)$, where $x\mapsto x^*$ denotes the \emph{Riesz duality isomorphism}) that is ortho-symplectic: $g_V(Bv,Bw)=g_V(v,w)$ and $\omega_V(B v,B w)=\omega_V(v,w)$, for all $v,w\in TM$. 

On $V:=TM$ we have an \emph{internal almost split-quaternionic structure} with generators $\id,B,R,F$:  
\begin{center}
$B^2=-\id, \quad R^2=+\id, \quad F:=B\circ R=-R\circ B$.  
\end{center}

\subsection[Classical Born Geometry and Split-quaternions $\slashed{\HH}$]{
Classical Born Geometry and Split-quaternions $\slashed{\HH}$ \cite{FLM14} \cite{BrGr10} \cite{NaZh25} \cite[section 3.5]{BC26}
}

A (flat) Born geometry is a generalized geometry $(V,\omega_V,J,\zeta_V)$, with $\RR$-structure $J$, equipped with a metric $g_V$ making $(V,g_V,\omega_V)$ K\"ahlerian (with polarizing $\CC$-structure $T$). 
Alternatively a Born geometry can be seen as a space equipped with an almost split-quaternionic structure and four bilinear forms $g_V$ (symmetric), $\omega_V$ (symplectic), $\eta_V$ and $\zeta_V$ (neutral symmetric) related as follows: 
\begin{align*}
\vcenter{\xymatrix{
  & \eta_V \ar@/^/[dd]^{-T}  & 
\\
\omega_V \ar@{<->}[ur]^{-L} \ar@{<->}[dr]_{J} \ar@/^/[rr]^{-T}   & & g_V \ar@/^/[ll]^{T} \ar@{<->}[ul]_{J}  \ar@{<->}[dl]^{L}
\\
 & \zeta_V \ar@/^/[uu]^{T} &   
}
}
\quad \quad \quad \quad 
\begin{aligned}
&\slashed{\HH}:=\Big\{\alpha I + \beta T + \gamma J + \delta L \ \big| \ \alpha,\beta,\gamma,\delta\in \RR \Big\}, 
\\
&\quad\quad \quad T^2=-I, \quad J^2=+I, \quad L^2=+I, 
\\
&[T,J]_+=[L,T]_+=[J,L]_+=0, \quad L\circ T\circ J=I, 
\end{aligned}
\end{align*} 
\begin{align*}
&g_V(+Tv,w)=\omega_V(v,w), && g_V(+Jv,w)=\eta_V(v,w), && g_V(+Lv,w)=\zeta_V(v,w), 
\\
&\omega_V(-Tv,w)=g_V(v,w), && \omega_V(+Jv,w)=\zeta_V(v,w), && \omega_V(-Lv,w)=\eta_V(v,w), 
\\ 
&\eta_V(-Tv,w)=\zeta_V(v,w), && \eta_V(+Jv,w)=g_V(v,w), && \eta_V(-Lv,w)=\omega_V(v,w), 
\\
&\zeta_V(+Tv,w)=\eta_V(v,w), && \zeta_V(+Jv,w)=\omega_V(v,w), && \zeta_V(+Lv,w)=g_V(v,w).  
\end{align*}
\begin{itemize}
\item[$\rightsquigarrow$] \it 
the previous structure can be compactly summarized saying that $V$ is a left-$\slashed{\HH}$-Hilbert \hbox{C*-module} over the $\RR$-C*-algebra $\slashed{\HH}$ of split-quaternions with the $\slashed{\HH}$-valued inner-product:  
\end{itemize}
\begin{equation*}
I\cdot g_V+ T\cdot \omega_V+ J\cdot \eta_V + L\cdot \zeta_V. 
\end{equation*}
We already have two fundamental examples of Born geometries coming from: 
\begin{itemize}
\item 
Born reciprocity of a polarized phase-space $V=X\oplus X^*$ with configuration-space $(X,g_X)$: 
\begin{equation*}
x\oplus y^*\xrightarrow{T:=B}(-y)\oplus x^*,  
\quad 
x\oplus y^*\xrightarrow{J:=R} x\oplus (-y^*),  
\quad 
x\oplus y^*\xrightarrow{L:=F}y\oplus x^*, 
\quad 
\forall x\in X,\ y^*\in X^*; 
\end{equation*}
\item 
standard $\RR$-subspaces $V_\RR$ of $\CC$-Hilbert spaces $\H_\CC:=\cj{V_\RR\oplus V_\RR}$ with unitary $\CC$-structure $T_V$ the multiplication by $i\in\CC$, anti-unitary modular conjugation as $\RR$-structure $J_V$ and $L_V:=T_V\circ J_V$. 
\end{itemize} 

\subsection[Modular Born Geometries and Hypercomplex Actions]{
Modular Born Geometries and Hypercomplex Actions \cite[section 3.3]{BC26}
}

The external complexification $\H_\CC$ of $V$ has a canonical unitary modular $V$-preserving flux: 
\begin{gather*}
\Delta_V^{it}=\exp(i K_V t), \quad K_V=\Gamma_V\circ\Theta_V, \quad \Theta_V:=|K_V|,  
\end{gather*}
where $\Gamma_V$ is the unitary phase of the polar decomposition of the modular generator $K_V$. 
\begin{itemize}
\item[$\rightsquigarrow$] 
$\Phi_V:=-T_V\circ \Gamma_V$ is an additional canonical {\it modular $\CC$-structure}\footnote{This complex structure on $V$ has been independently identified also in \cite[corollary 1.6.6]{CLL26}.} on $\H_\CC$: $\Phi_V^2=-I_V$, 
\item[$\rightsquigarrow$] 
$\Phi_V$ is a canonical \emph{modular Born symmetry} on $\H_\CC$: it coincides with the phase of the polar decomposition of the complexification of the polarizers\footnote{
We further assume here (and for now on) the invertibility of $P$. 
} 
$P, Q$ of $(V,g_V,\omega_V)$: 
\begin{gather*}
\Phi_V=B\oplus B, \quad 
\omega_V(v,w)=g_V(Pv,w), \quad g_V(v,w)=\omega_V(v,Qw), \quad \forall v,w\in V, 
\\
P=B\circ |P|, \quad Q=B\circ|Q|, \quad |Q|=|P|^{-1}, 
\end{gather*}
\end{itemize}
The operators $J_V, \Phi_V, T_V$ generate, on $\H_\CC\simeq\CC\otimes_\RR V$, the hypercomplex $\RR$-algebra of the $q$-bit $\MM_2(\CC)_\RR\simeq\CC_\RR\otimes_\RR\slashed{\HH}$, a $\Phi_V$-complexification of the split-quaternions $\slashed{\HH}$ generated by $T_V,J_V,L_V$. 

We will see that the ``first-quantized thermal-double'' $V_\RR\oplus V'_\RR$ has a {\it quantized} Born geometry.  

\subsection[Modular CPT-symmetry and Modular $\CC$-polarization]{
Modular CPT-symmetry and Modular $\CC$-polarization \cite[pages 22 and 25]{BC26}
}

Inside the external complexification $(\H_\CC,h)$ of $V$ we have the following symmetries relating $V$ with its symplectic complement $V'$ and its metric complement $V^\perp$:
\begin{align*}
\begin{aligned}
&V^\perp:=
\{v\oplus w  | \forall z, \re h (v\oplus w,z\oplus 0)=0\},
\\ 
&V'=J_VV\simeq\cj{V}, \quad V^\perp=L_V V\simeq V^*, 
\\  
&(V')^\perp=(V^\perp)'=T_VV=:iV, 
\end{aligned}
\quad \quad \quad \quad \quad \quad \quad  
\begin{aligned}
\xymatrix{
V' \ar@{<.>}[rr]|{J_V} \ar@{<.>}[d]|{L_V} \ar@{<.>}[rrd]
& & V \ar@{<.>}[d]|{L_V}
\\
iV \ar@{<.>}[rr]|{J_V} \ar@{<.>}[rru]|{T_V} & & V^\perp
}
\end{aligned}
\end{align*}
\begin{itemize}
\item[$\rightsquigarrow$] 
the modular conjugation $J_V$ can be interpreted as a canonical \emph{modular CPT-symmetry}, 
\item[$\rightsquigarrow$]  
the modular phase $\Gamma_V$ of the modular generator $K_V$ is a \emph{modular $\CC$-polarization}: $\Gamma_V^2=I_V$, 
\item[$\rightsquigarrow$] 
the $\Gamma_V$-eigenspaces $\Gamma_V^\pm$ are modularly $\Delta_V^{it}$-stable, 
and describe (see the next section) a pair of {\it internally $\pm B$-complexified CPT-mirror copies} of $(V,g_V,\omega_V)$: 
\begin{equation*}
\Gamma^+_V=\{v\oplus Bv \ | \ v\in V\}, 
\quad  
\Gamma^-_V=\{Bv\oplus v \ | \ v\in V\}, 
\quad \quad 
\begin{aligned}
\xymatrix{
\Gamma^+_V \ar@{<.>}[rr]|{J_V} \ar@{<.>}[d]|{L_V} \ar@{<.>}[rrd]
& & \Gamma^-_V \ar@{<.>}[d]|{L_V}
\\
\Gamma^-_V \ar@{<.>}[rr]|{J_V} \ar@{<.>}[rru]|{T_V} & & \Gamma^+_V
}
\end{aligned}  
\end{equation*}
\item[$\rightsquigarrow$]  
we conjecture that $L_V$ plays the role of an \emph{external duality operator} $L_V(\Gamma^-_V)=\Gamma^+_V\simeq (\Gamma^-_V)^*$.
\end{itemize}

\subsection[Internal Complexifications and Mauro-Weyl Quantization]{
Internal Complexifications and Mauro-Weyl Quantization \cite[sections 3.2-3.3]{BC26}
}

A $\CC$-structure $B$ on an ortho-symplectic space $(V,g_V,\omega_V)$, induces an \emph{internal complexification} $\K_\CC^B$ with inner product given, for all $v,w\in V$, by:  
\begin{multline*}
k_{B_V}(v,w):=
\frac{1}{4}
\big(
+g_V(v,w) +\omega_V(v,Bw) 
-\omega_V(Bv,w) +g_V(Bv,Bw)
\big)
\\
+\frac{i}{4}\cdot
\big(
+\omega_V(v,w) -g_V(v,Bw) 
+g_V(Bv,w) +\omega_V(Bv,Bw)   
\big),
\end{multline*}
that for the special case of ortho-symplectic $\CC$-structures takes the form: 
\begin{equation*}
k_{B_V}(v,w)=\frac{1}{2}\cdot\Big(g_V(v,w) + \omega_V(v,Bw) \Big)+\frac{i}{2}\cdot\Big(\omega_V(v,w) + g_V(Bv,w)\Big), 
\end{equation*}
that in the K\"ahlerian case reproduces: $k_{B_V}(v,w)=g_V(v,w)+i\cdot\omega_V(v,w)$.

Considering the two $\CC$-eigenspaces $\Gamma^\mp_V$ of the $\CC$-polarization $\Gamma_V$ on $\H_\CC\simeq\Gamma^+_V\oplus\Gamma^-_V$, we have:  
\begin{itemize}
\item[$\rightsquigarrow$] 
\emph{$\Gamma^\mp_V$ is the internal $(\pm B)$-complexification of $(V,\frac{1}{2}g_V,\frac{1}{2}\omega_V)$, 
\item[$\rightsquigarrow$] 
$v\mapsto v\oplus \pm B v$, for $v\in V$, is an algebraic D.Mauro first-quantization map \cite{Ma03}, 
\item[$\rightsquigarrow$] 
Weyl second-quantization \cite[section 4]{BGP07} produces a CPT-thermal-double 
\begin{gather*}
\WWy(\Gamma_V^+)\otimes_\CC\WWy(\Gamma_V^-), \quad \quad  \WWy(\Gamma_V^+)\simeq \WWy(\Gamma_V^-)'
\quad \quad 
\xymatrix{\WWy(\Gamma_V^+) \ar@{<->}[rr]^{\WWy(J_V)} & & \WWy(\Gamma_V^-)}. 
\end{gather*}
}
\end{itemize}
 
\subsection[Modular Flux and Modular Inverse Temperature $\Theta_V$]{
Modular Flux and Modular Inverse Temperature $\Theta_V$ \cite[pages 22-25]{BC26}
}

For a $\RR$-structure $S_V$ on a $\CC$-Hilbert space $\H_\CC$ (with its $\CC$-structure $T_V\xi=i\cdot\xi$) we have: 
\begin{gather*}
\left[\text{$T_V$ and $S_V$ generate $\slashed{\HH}$}\right]
\iff
\left[\text{$V\subset\H_\CC$ standard $\RR$-subspace of $\H_\CC$}\right],
\\
\left[\text{$S_V$ not anti-unitary on $\H_\CC$}\right]
\iff
\left[\text{non-trivial\phantom{$|$}modular flux 
}\right], 
\end{gather*}

\begin{itemize}
\item[$\rightsquigarrow$]
K\"ahler spaces $(V,g_V,\omega_V)$ have divergent modular flux (and semi-Hilbert complexifications),  
\item[$\rightsquigarrow$] 
for (non-K\"ahlerian) ortho-symplectic spaces we have the equivalences: 
\begin{equation*}
\boxed{\text{
\begin{minipage}{2.0cm} 
\begin{center}
non-trivial modular flux  
\end{center}
\end{minipage}
}}
\iff
\begin{aligned}
\boxed{\xymatrix{
iV\neq V^\perp
\ar@{<=>}[d]
\\
\re h|_{V\times iV}\neq 0
}}
\end{aligned}
\iff
\begin{aligned}
\boxed{
\xymatrix{
V\neq V' 
\ar@{<=>}[d]
\\
\im h|_V\neq 0
}}
\end{aligned}
\iff 
\boxed{\text{
\begin{minipage}{2.8cm} 
\begin{center}
non-commutative $V\subset\WWy(V)$  
\end{center}
\end{minipage}
}}
\end{equation*} 
\item[$\rightsquigarrow$] 
invertible modular inverse temperature $\Theta_V$ $\iff P$ invertible $\iff$ \emph{fully} non-trivial modular flux: 
\begin{gather*}
\Theta_V:=|K_V|=\begin{bmatrix} \hat{\theta} & 0 \\ 0 & \hat{\theta} \end{bmatrix}, 
\quad  
\hat{\theta}:= \log \left(\frac{\id_V+|P|}{\id_V-|P|}\right), 
\quad   
\Delta^{it}_V=\exp(-t\cdot \Phi_V\circ\Theta_V). 
\end{gather*}
\end{itemize}

\subsection[Modular Born $\RR$-polarizations and Non-commutative-spaces]{
Modular Born $\RR$-polarizations and Non-commutative-spaces \cite[pages 26-27]{BC26}
}

We couldn't find a canonical Born $\RR$-polarization induced by modular theory! 
As a consequence we take seriously \emph{Born relative localization} proposal \cite{A-CFK-GS11a,A-CFK-GS11b} and we define 
\begin{itemize}
\item[$\rightsquigarrow$] 
a \emph{modular Born $\RR$-polarization} as any $\RR$-structure $\Lambda_V$ on $\H_\CC$, commuting with $\Gamma_V$, and generating with the modular $\CC$-polarization $\Phi_V$ and interal copy of the split-quaternions 
$\slashed{\HH}$: 
\begin{equation*}
\Lambda^2_V=I_V, \quad [\Lambda_V,\Phi_V]_+=0, \quad [\Lambda_V,\Gamma_V]_-=0.
\end{equation*} 
\end{itemize}

Each choice of $\Lambda_V$ produces standard $\RR$-eigen-subspaces of ``coordinates'' 
$\Lambda_V^+\cap \Gamma_V^\pm\subset\Gamma_V^\pm$ and of ``momenta'' $\Lambda_V^-\cap \Gamma_V^\pm\subset\Gamma_V^\pm$ in both the $J_V$-conjugated one-particle Hilbert spaces $\Gamma_V^\mp$. 

Generically our Born $\RR$-polarization $\Lambda_V$ will have a non-trival modular flux\footnote{
Trivial fluxes on some proper subspace can happen if $\Theta_{\Lambda^\pm_V}$ is not invertible.
}
and will reduce to a ``classical'' fully degenerate polarization only when the modular flux is fully trivial ($\Theta_{\Lambda_V^\pm}=0$): 
\begin{equation*} 
\big[\text{quantum coordinates/momenta spaces $\Lambda^\pm_V$ }\big]
\iff  
\big[\text{$\Lambda_V$ has non-trivial modular flux:  $\Theta_{\Lambda_V^\pm}\neq 0$}\big]. 
\end{equation*}

\subsection[Modular Quantum Born Geometries]{
Modular Quantum Born Geometries \cite[page 26 and section 3.5]{BC26}
}

Consider a Hitchin-Gualtieri generalized geometry $(V:=X\oplus X^*,\omega_V,R,\zeta_V)$ with 
\begin{gather*}
\omega_V(x_1\oplus y_1^*,x_2\oplus y_2^*)=y_2^*(x_1)-y_1^*(x_2), 
\quad  
\zeta_V(x_1\oplus y_1^*,x_2\oplus y_2^*)=y_2^*(x_1)+y_1^*(x_2), 
\\
x\oplus y^* \xrightarrow{R} x\oplus(-y^*), \quad 
\forall x_1,x_2,x\in X, \ y_1^*,y_2^*,y^*\in X^*. 
\end{gather*}
Introducing a metric $g_V$ that, in the polarization $R$, is ``diagonal'' 
$g_V(Rv,Rw)=g_V(v,w), \ v,w\in V$, 
produces a \emph{modular deformation of a classical Born geometry $(V,\omega_V,R,\xi_V,g_B)$ of $V$}: 
\begin{gather*}
\begin{aligned}
\xymatrix{
&  \eta_V \ar@/^/[dl]|{-F} \ar@/^/[dr]|{+R}  &  
\\
\omega_V \ar@/^/[rr]|{\ -B} \ar@/^/[dr]|{+R} \ar@/^/[ur]|{-F} 
& & 
g_{B} \ar@/^/[ll]|{\ +B} \ar@/^/[dl]|{+F} \ar@/^/[ul]|{+R} 
\\ 
&  \ar@/^/[ul]|{+R} \zeta_V \ar@/^/[ur]|{+F} &  
}
\end{aligned}
\
\rightsquigarrow
\ 
\begin{aligned}
\xymatrix{
& & \eta_V  \ar@/^/[dll]|{-F} \ar@/^/[drr]|{\quad \quad +R\circ |Q|} & & 
\\
\omega_V \ar@/^/[rrrr]|{\ -Q} \ar@/^/[drr]|{+R} \ar@/^/[urr]|{-F}
& & & & g_V \ar@/^/[llll]|{\ +P} \ar@/^/[dll]|{\quad \quad +|P|\circ F}  \ar@/^/[ull]|{+|P|\circ R \quad \quad } 
\\ 
& & \ar@/^/[ull]|{+R} \zeta_V \ar@/^/[urr]|{+F\circ |Q|\quad \quad } & & 
}
\end{aligned} 
\\ 
P=B\circ |P|, 
\quad   
|P|= \frac{\cosh(\hat{\theta})-\id_V}{\sinh(\hat{\theta})}=|Q|^{-1}, 
\quad  
\hat{\theta}=\Theta_V|_V,
\quad  
g_V(v,w)=g_B(|P|v,w), \quad v,w\in V. 
\end{gather*}
\begin{itemize}
\item[$\rightsquigarrow$] 
``off-diagonal'' metrics $g_V$ will induce {\it Kalb-Ramond deformations} 
(see B.Jur\v co, P.Schupp, J.Vysoky \cite{JSV13}) that are related to the non-trivial modular theories of the no-more-orthogonal polarizations $R$ and $F$: in this case coordinates $X$ and momenta $X^*$ are non-commutative. 
\end{itemize}

\section[Extremely Conjectural Developments and Possible Physical Implications]
{Extremely Conjectural Developments and Possible Physical Implications}

We summarize here some still conjectural (physical) developments of the present formalism. 

\subsection[Tulczyjew Double Bundles and External Born Reciprocity]{
Tulczyjew Double Bundles and External Born Reciprocity \cite{Tul74} \cite{GrGr22} \cite[section 3.1]{BC26} 
}

All the material presented up to this moment is essentially of a local (point-wise) nature applying to each fiber of the bundle $\CC\otimes_\RR (TX\oplus T^*X)$, over a Riemannian manifold $(X,g_X)$. 

Globally, we suggest to consider a \emph{non-linear momentum manifold} $\hat{X}\simeq X$ diffeomorphic to $X$, equipped with a \emph{Riesz dual metric} $g_{\hat{X}}$ such that, for all $p\in X$, $(T_{\hat{p}}\hat{X},g_{\hat{X}})\simeq (T_{p}^*X,g^*_{X})$, where $g^*_X$ denotes the metric induced on $T_{p}^*X$ by the Riesz isomorphism $G^X_p:T_pX\to T_p^*X$. 

In practice we have the following commuting diagram of Tulczyjew double bundles: 
\begin{equation*}
\xymatrix{\ar[d]
T(X)\oplus_X T^*(X) \simeq T(T^*X) \ar@{->}[rrrrr]^{\begin{minipage}{4.7cm}{\begin{center}\footnotesize ``Tulczyjew'' 
duality isomorphism $\tau_X$\end{center}}\end{minipage}\quad}_{T^*\delta_X\circ T^*G^X\circ\tau_X=\tau_{\hat{X}}\circ TG^{\hat{X}}\circ T\delta_X} & & & & &
T^*(T\hat{X}) \simeq T^*(\hat{X})\oplus_{\hat{X}} T^*(\hat{X}) 
\ar[d] 
\\ 
(T^*(X), g^*_{X}) \ar@{->}[rrrrr]^{\begin{minipage}{5cm}{\begin{center}\footnotesize Riesz duality isometric isomorphism \end{center}}\end{minipage}}_{\delta_X}
\ar[d] 
& & & & & (T(\hat{X}),g_{\hat{X}}) \ar[d]
\\
(X,g_X) \ar[rrrrr]^{\begin{minipage}{5cm}{\begin{center}\footnotesize non-linear duality diffeomorphism \end{center}}\end{minipage}}_{x\mapsto\hat{x}} & & & & & (\hat{X},g_{\hat{X}}). 
}
\end{equation*} 
In this ``enlarged setting'' Born reciprocity remains globally defined at the non-linear level treating coordinates $X$ and momenta $\hat{X}$ in perfectly symmetric way. 
\begin{itemize}
\item[$\rightsquigarrow$] \it 
there are dual modular flows $\sigma^{g_X,\gamma}$ and $\sigma^{g_{\hat{X}}\hat{\gamma}}$ acting as a dynamics constraints on the complexified of the Tulczyjew bundles of coordinates $X$ and momenta $\hat{X}$. 
\end{itemize}

The full theory (involving relative modular operators) seems to be based on a complexified generalized bundle over $X\times\hat{X}$ and it is a natural to ponder the existence of \emph{global non-linear modular flows} on $X\times\hat{X}$. This will be left to future investigations. 

\subsection[CPT-mirror Quantum Born Geometries and Cayley-Dixon Products]{
CPT-mirror Quantum Born Geometries and Cayley-Dixon Products \cite[pages 32-33]{BC26} 
}
 
Upon external complexification and $\CC$-polarization, we obtain a quantum version of the classical CPT-Born double, with first-quantized $\CC$-Hilbert spaces $\Gamma^\pm$ and $(\Gamma^{\pm})^*$ related via the action of an external copy of split-quaternions $\slashed{\HH}$ generated by the CPT symmetry $J$ and the Born duality $L$: 
\begin{equation*}
\xymatrix{
\boxed{\begin{minipage}{5.0cm}
\begin{center}
$(\hat{X}',\hat{g}'_X,\hat{\omega}'_X)$ \quad {$(\Gamma^+_V)^*$}
\\
CPT-mirror momentum space 
\end{center}
\end{minipage}}
& & & \ar@{<.>}[lll]_{\quad \text{CPT-symmetry}}^{\quad J}
\boxed{\begin{minipage}{3.8cm}
\begin{center}
$(\hat{X},\hat{g}_{X},\hat{\omega}_{X})$ \quad {$(\Gamma^-_V)^*$}
\\
momentum space
\end{center}
\end{minipage}}
\\ 
\boxed{\begin{minipage}{5.0cm}
\begin{center}
$(X',g'_{X},\omega'_{X})$ \quad {$\Gamma^+_V$}
\\
CPT-mirror configuration space 
\end{center}
\end{minipage}}
\ar@{<.>}[u]|{\text{external Born duality} \ L} 
& & & \ar@{<.>}[u]|{\text{external Born duality} \ L}
\boxed{\begin{minipage}{3.8cm}
\begin{center}
$(X,g_X,\omega_X)$ \quad {$\Gamma^-_V$}
\\
configuration space 
\end{center}
\end{minipage}}
\ar@{<.>}[lll]_{\quad \text{CPT-symmetry}}^{\quad J}
}
\end{equation*}
Once a modular Born $\RR$-polarization $\Lambda$ has been selected, the holomorphic and anti-holomorphic Hilbert first-quantization spaces $\Gamma_V^\mp$ both acquire a further internal structure of $\slashed{\HH}$-Hilbert \hbox{C*-mod}\-ules over the non-commutative $\RR$-C*-algebra $\slashed{\HH}$ of split-quaternions (internal Born reciprocity). 

For such Hilbert modules over non-commutative algebras, conjugates and duals are structurally different spaces, so the identifications $(\Gamma_V^\mp)^*\simeq\Gamma_V^\pm$ seem suspicious, and hence it seems natural to conjecture that the classical symmetries $\slashed{\CC}\otimes_\RR\slashed{\HH}$ (respectively $\slashed{\HH}\otimes_\RR\slashed{\HH}$ in the polarized case) of the external complexification of $(V,g_V,\omega_V)$ might be replaced by split-octonionic $\slashed{\CC} \bowtie \slashed{\HH}$ (respectively split-sedenionic $\slashed{\HH}\bowtie\slashed{\HH}$ in the polarized case), where here $\bowtie$ denotes the {\it Cayley-Dixon product}. 

In this case, the centrality of the modular Born symmetry $\Phi_V$ and the associativity in the hypercomplex algebra $\MM_2(\CC)_\RR $ might just be artefacts of a ``$\CC$-Hilbert space projection'' of a split-octonionic structure originating from a new modular theory on a split-quaternionification of $V$. 
 
\subsection[Modular Dynamics and Geometric Pre-quantization]{
Modular Dynamics and Geometric Pre-quantization \cite[section 3.6]{BC26}
}

The fundamental idea that global Born reciprocity establishes a dual correspondence between 
{\footnotesize
\begin{equation*}
\xymatrix{
\boxed{
\begin{minipage}{6cm} 
\begin{center}
gravity 
\\
curvature of coordinates 
\\ 
non-commutative momenta 
\end{center}
\end{minipage}
} 
\ar@{<.>}[rr]^{\text{Born reciprocity}} & & 
\boxed{
\begin{minipage}{6cm} 
\begin{center}
co-gravity \\ 
curvature of momenta 
\\
non-commutative coordinates 
\end{center}
\end{minipage}
} 
}
\end{equation*}
}
originates from S.Majid's work on the {\it principle of representation-theoretic self-duality} \cite{Maj88,Maj91}; in our case (after a global modular $\RR$-polarization $\Lambda$ has been selected) the Born-reciprocal non-commutative geometries of coordinates/momenta are of modular origin and are induced by dual modular Abelian gauge fluxes as described, for arbitrary Abelian gauge theories, by E.Guendelman, D.Singleton \cite{GuSi22}, with modular temperatures as Born reciprocal gauge potentials. 

One might use \emph{modular time} to construct an associated $\CC$-line-bundle on phase-space and \emph{modular inverse temperature} to produce a connection on such line-bundle,
as long as a metric is fixed, this should provide a \emph{canonical geometric pre-quantization} of a classical theory (and also suggests the possibility of developing a \emph{thermodynamics treatment of individual quantum systems}).  

The modular quantum dynamics (for a second-quantized system) induces a Weinstein \cite{Wei97} classical modular (harmonic oscillator) dynamics that, under suitable limits, ``converges'' to the geodesic Liouvillian flow on phase-space.  

The periodic modular group of a Riemannian configuration space $(X,g_X)$ provides an explaination for the origin of the $U(1)$-central extensions that characterize the ``transition'' between the classical and the quantum dynamical groups (see \cite{FRS13} and references therein): 
{\footnotesize
\begin{center} 
$\xymatrix{
i\cdot \RR \ar[rr] & & 
\boxed{\text{Poissonian morphisms}} 
\ar[rr]
& & 
\boxed{
\text{Hamiltonian morphisms}}
}$ 
\\ 
$\xymatrix{
U(1)
\ar[rr] & & 
\boxed{
\text{Quanto-morphisms}} 
\ar[rr]
& & 
\boxed{\text{Hamiltonian morphisms}}
}$. 
\end{center}
}

\subsection[Tomita CPT, Contravariant Non-commutative Geometry, Mirror Cosmology]{
Tomita CPT, Contravariant NCG, Mirror Cosmology \cite[page 38 and 40]{BC26}
}

A basic aspect of modular algebraic quantum theory is that \emph{CPT-symmetry} is a built-in feature of local \emph{reducible} standard GNS-representations, of a modular local algebra $\As$, induced by a covariant vacuum state $\phi$: they are representations of \emph{thermal-doubles} $\As\otimes (J_\phi\As J_\phi)$.  
The physical relevance of these thermal-representations has been known for quite a while \cite{Str66} \cite{Wor72} as well as their relation with (pre-)quantization, modular theory and thermo-field dynamics (see for example \cite{Is76} \cite{Emc81,Oj81}). 

\medskip

Such ``chiral symmetry'' seems also a direct byproduct of non-commutativity of space-time: non-commutative vector fields (derivations) and connections split in right/left variants (see for example \cite{Bor97}, \cite{VdB04} and also \cite{Pu21}, where a program for a {\it contravariant version of non-commutative differential geometry}, making explicit usage of modular thermal-doubles, is put-forward). 

\medskip

A.Sakharov \cite{Sa67,Sa80} (originally attempting to resolve the problems of baryogenesis, matter-antimatter and time asymmetry) and more recently L.Boyle, K.Finn and N.Turok \cite{BFT18a,BFT18b,BT21} (addressing some
of the pressing issues with the current $\Lambda$-CDM inflationary standard cosmological model) have suggested alternative semi-classical \emph{CPT-symmetric cosmologies}. 
\begin{itemize}
\item[$\rightsquigarrow$] \it
``modular algebraic quantization'' can provide a quantum-level justification (with a modular origin) for these CPT-mirror cosmologies: every local modular algebra of covariant-observables in our ``matter'' universe has a CPT-mirror copy to be interpreted as a local modular algebra based on the corresponding region of a CPT-mirror ``anti-matter''universe. 
\end{itemize}
The formalism here presented also allows more unconventional revisitations of (CPT-mirrored versions) of (now officially dismissed) steady-state cosmologies, where the CMB-radiation might originate from a thermal gravitational-horizon effect of the modular thermal-double Born geometry. 

\subsection[Possible Relation with Interferometry Experiments]{
Possible Relation with Interferometry Experiments \cite[pages 38-39]{BC26}
} 

Some theoretical developments, broadly inspired by (variations of) AdS/CFT (celestial) holography, claim that: 
the quantum nature of the geometry induced by modular theory might actually be soon testable, for example by the \href{https://magazine.caltech.edu/post/quantum-gravity}{GQuEST} collaboration proposed interferometry experiments based on the works by K.Zurek, E.Verlinde \cite{VeZu19a,VeZu19b,VeZu22}.\footnote{
I became aware of this very intriguing ongoing experimental attempts at the QG 2023 Nijmegen \href{https://indico.imapp.ru.nl/event/106/contributions/298/attachments/157/311/23_QG_Zurek\%20Kathryn\%20Zurek\%20\%281\%29.pdf}{K.Zurek's talk}. 
}

The purpose of such experiments is to observe the quantum superposition/fluctuation nature of the geometry induced by the vacuum state of (conformal) free quantum scalar field theory on Minkowski spacetime via the modular theory of double cones algebras (seen as conformal equivalent to black holes). 
\begin{itemize}
\item[$\rightsquigarrow$]
positive results of GQuEST will support a direct ``quantum geometrical'' meaning of modular theory, as suggested in this modular algebraic quantum gravity research program. 
\end{itemize}

\appendix 

\acknowledgments
We hereby acknowledge a research visit to Warsaw in 2024 that was supported by the University of Warsaw Thematic Research Programme “Quantum Symmetries”. 

PB is grateful to Starbucks Coffee (Emsphere Tower, Langsuan, UBCII Tower, One Bangkok, Gaysorn Village, Emquartier
Sky Garden, Central World Office Tower and Jasmine City) where he spent most of the time dedicated to this research
project. 

\medskip 

\noindent
{\bf Declarations}
 
This is a ``pre-AI-era'' work and no direct use of LLMs, or other artificial intelligence aid, has been made during its conceptual elaboration and its writing. 



\begin{thebibliography}{99}

\bibitem{A-CFK-GS11a}

G. Amelino-Camelia, L. Freidel, J. Kowalski-Glikman, L. Smolin, 
\emph{The Principle of Relative Locality}, 
\emph{\href{https://doi.org/10.1103/PhysRevD.84.084010}{Phys.~Rev.~D}} {\bf 84} (2011) 084010 
[{\tt \href{https://arxiv.org/abs/1101.0931}{arXiv:1101.0931}}]. 

\bibitem{A-CFK-GS11b}

G. Amelino-Camelia, L. Freidel, J. Kowalski-Glikman, L. Smolin,  
\emph{Relative Locality: A Deepening of the Relativity Principle},  
\emph{\href{https://doi.org/10.1007/s10714-011-1212-8}{Gen.~Relativ.~Gravit.}} {\bf 43}(10) (2011) 2547-2553 
[{\tt \href{https://arxiv.org/abs/1106.0313}{arXiv:1106.0313}}].

\bibitem{Ar63}

H. Araki, 
\emph{A Lattice of Von Neumann Algebras Associated with the Quantum Theory of a Free Bose Field}, 
\emph{\href{https://doi.org/10.1063/1.1703912}{J.~Math.~Phys.}} {\bf 4} (1963) 1343-1362. 

\bibitem{Ar64}

H. Araki, 
\emph{Von Neumann Algebras of Local Observables for Free Scalar Field}, 
\emph{\href{https://doi.org/10.1063/1.1704063}{J.~Math.~Phys.}} {\bf 5} (1964) 1-13. 

\bibitem{BGP07}

C. B\"ar, N. Ginoux, F. Pf\"affle, 
\emph{Wave Equations on Lorentzian Manifolds and Quantization}, 
\emph{\href{https://doi.org/10.4171/037}{ESI Lectures in Mathematical Physics}} 
European Mathematical Society (EMS) 2007 
[{\tt \href{https://arxiv.org/abs/0806.1036}{arXiv:0806.1036}}].

\bibitem{B23}

P. Bertozzini, 
\emph{\href{https://doi.org/10.13140/RG.2.2.33754.71367}{Modular Algebraic Quantum Geometries for Quantum Gravity}}, 
BIRS workshop ``Gravity, Noncommutative Geometry, Cosmology'', Oaxaca, Mexico, 25 August 2023.  

\bibitem{BCL10}

P. Bertozzini, R. Conti, W. Lewkeeratiyutkul 
\emph{Modular Theory, Non-commutative Geometry and  Quantum Gravity}, 
\href{https://doi.org/10.3842/SIGMA.2010.067}{
\emph{SIGMA} {\bf 6} (2010) 067} 
[{\tt \href{https://arxiv.org/abs/1007.4094v2}{arXiv:1007.4094v1}}]. 

\bibitem{BC26}
P. Bertozzini, R. Conti,  
\emph{Modular Quantization and Non-commutative Space-time}, 
\newline 
[{\tt \href{https://www.alphaxiv.org/abs/2611.modular-quantization-noncommutative-spacetime}{alphaXiv:2611.modular-quantization-noncommutative-spacetime}}].
 
\bibitem{Bo38}

M. Born,
\emph{A Suggestion for Unifying Quantum Theory and Relativity}, 
\emph{\href{https://doi.org/10.1098/rspa.1938.0060}{Proc.~R.~Soc.~A}} {\bf 165}(921) (1938) 291-303.

\bibitem{Bor97}

A. Borowiec, 
\emph{Vector Fields and Differential Operators: Noncommutative Case}, 
\emph{\href{https://doi.org/10.1023/A:1021697831180}{Czech.~J.~Phys.}} {\bf 47}(11) (1997) 1093-1100
[{\tt \href{https://arxiv.org/abs/q-alg/9710006}{arXiv:q-alg/9710006}}]. 

\bibitem{BFT18a}

L. Boyle, K. Finn, N. Turok, 
\emph{CPT-Symmetric Universe}, 
\emph{\href{https://doi.org/10.1103/PhysRevLett.121.251301}{Phys.~Rev.~Lett.}} {\bf 121}(25) (2018) 251301 
[{\tt \href{https://arxiv.org/abs/1803.08928}{arXiv:1803.08928}}]. 

\bibitem{BFT18b}

L. Boyle, K. Finn, N. Turok, 
\emph{The Big Bang, CPT, and Neutrino Dark Matter}, 
\emph{\href{https://doi.org/10.1016/j.aop.2022.168767}{Annals Physics}} {\bf 438} (2022) 168767
[{\tt\href{https://arxiv.org/abs/1803.08930}{arXiv:1803.08930}}]. 

\bibitem{BT21}

L. Boyle, N. Turok, 
\emph{Two-Sheeted Universe, Analyticity and the Arrow of Time}, 
\newline 
[{\tt \href{https://arxiv.org/abs/2109.06204}{arXiv:2109.06204}}] 2021. 

\bibitem{BrGr10}

D. C. Brody, E.-M. Graefe,
\emph{On Complexified Mechanics and Coquaternions}, 
\emph{\href{https://doi.org/10.1088/1751-8113/44/7/072001}{J.~Phys.~A: Math.~Theor.}} {\bf 44}(7) (2011) 072001
[{\tt \href{https://arxiv.org/abs/1012.0757}{arXiv:1012.0757}}].

\bibitem{Ca21}

H. Capellmann,
\emph{Space-Time in Quantum Theory}, 
\emph{\href{https://doi.org/10.1007/s10701-021-00441-0}{Found.~Phys.}} {\bf 51} (2021) 44 
\newline 
[{\tt \href{https://doi.org/10.1007/s10701-021-00441-0}{arXiv:2006.00503}}].

\bibitem{Cic09}

N. Ciccoli, 
\emph{The Modular Class and Its Quantization - Minicourse}, 
\emph{\href{https://www2.im.uj.edu.pl/actamath/PDF/47-29-50.pdf}{Acta Math.~Univ.~Jag.}} {\bf 47} (2009) 29-50. 

\bibitem{CoRo94}

A. Connes, C. Rovelli, 
\emph{Von Neumann Algebra Automorphisms and Time-thermodynamics Relation in General Covariant Quantum Theories}, 
\emph{\href{https://doi.org/10.1088/0264-9381/11/12/007}{Class.~Quantum Gravity}} {\bf 11} (1994) 2899–2917
[{\tt \href{https://arxiv.org/abs/gr-qc/9406019}{arXiv:gr-qc/9406019}}].

\bibitem{CLL26}

R. Correa Da Silva, G. Lechner, R. Longo, 
\emph{Standard Subspaces and Modular Theory General Theory and Applications to QFT}, 
[{\tt \href{https://arxiv.org/abs/2611.08668v1}{arXiv:2611.08668v1}}] 2026. 

\bibitem{Emc81}

G. G. Emch, 
\emph{Prequantization and KMS Structures} 
\emph{\href{https://doi.org/10.1007/bf00670552}{Int.~J.~Theor.~Phys.}} {\bf 20}(12) (1981) 891-904.

\bibitem{DrWa21}

N. Drago, S. Waldmann, 
\emph{Classical KMS Functionals and Phase Transitions in Poisson Geometry}, 
\emph{\href{https://doi.org/10.4310/jsg.2023.v21.n5.a3}{J.~Symplectic Geom.}} {\bf 21}(5) (2023) 939-995
[{\tt \href{https://arxiv.org/abs/2107.04399}{arXiv:2107.04399}}]. 

\bibitem{FRS13}

D. Fiorenza, C. L. Rogers, U. Schreiber,
\emph{Higher U(1)-gerbe Connections in Geometric Prequantization}
\emph{\href{https://doi.org/10.1142/s0129055x16500124}{Reviews in Mathematical Physics}} {\bf 28}(6) (2016) 1650012
[{\tt \href{https://arxiv.org/abs/1304.0236}{arXiv:1304.0236}}].

\bibitem{FLM14}

L. Freidel, R. G. Leigh, D. Minic,  
\emph{Born Reciprocity in String Theory and the Nature of Spacetime}, 
\emph{\href{https://doi.org/10.1016/j.physletb.2014.01.067}{Phys.~Lett.~B}} {\bf 730} (2014) 302 
[{\tt \href{https://arxiv.org/abs/1307.7080}{arXiv:1307.7080}}]. 

\bibitem{GrGr22}

K. Grabowska, J. Grabowski, 
\emph{Contact Geometric Mechanics: the Tulczyjew Triples},
\emph{\href{https://doi.org/10.4310/atmp.240914022224}{Adv.~Theor.~Math.~Phys.}} {\bf 28}(2) (2024) 599-654 [{\tt \href{https://arxiv.org/abs/2209.03154}{arXiv:2209.03154}}].

\bibitem{Gu04}

M. Gualtieri, 
\emph{Generalized Complex Geometry},
D.Phil.~Thesis, Oxford University 2004
[{\tt\href{https://arxiv.org/abs/math/0401221}{arXiv:math/0401221}}],  
\emph{\href{https://doi.org/10.4007/annals.2011.174.1.3}{Ann.~Math.}} {\bf 174}(1) (2011) 75-123 
[{\tt \href{https://arxiv.org/abs/math/0703298}{arXiv:math/0703298}}].

\bibitem{GuSi22}

E. Guendelman, D. Singleton D, 
\emph{Momentum Gauge Fields and Non-Commutative Space-Time}, 
\emph{\href{https://doi.org/10.3390/sym15010126}{Symmetry}} {\bf 15} (1) (2023) 126 
[{\tt \href{https://arxiv.org/abs/2206.02638}{arXiv:2206.02638}}].

\bibitem{Hi10}

N. Hitchin, 
\emph{Lectures on Generalized Geometry}, 
Conan Leung  N, Yau S-T (eds)
\emph{\href{https://doi.org/10.4310/sdg.2011.v16.n1.a3}{Surv.~Differ.~Geom.}} {\bf XVI} (2011) 79-124 
[{\tt \href{https://arxiv.org/abs/1008.0973}{arXiv:1008.0973}}].

\bibitem{Is76}

W. Israel, 
\emph{Thermo Field Dynamics of Black Holes}, 
\emph{\href{https://doi.org/10.1016/0375-9601(76)90178-x}{Phys.~Lett.~A}} {\bf 57} (1976) 107-110. 

\bibitem{JSV13} 

B. Jurco, P. Schupp, J. Vysoky, 
\emph{On the Generalized Geometry Origin of Noncommutative Gauge Theory}, 
\emph{\href{https://doi.org/10.1007/jhep07(2013)126}{J.~High Energy Phys.}} {\bf 7} (2013) 126
[{\tt \href{https://arxiv.org/abs/1303.6096}{arXiv:1303.6096}}]. 

\bibitem{Lo08}

R. Longo,
\emph{\href{http://www.mat.uniroma2.it/~longo/Lecture\_Notes.html}{Lectures on Conformal Nets}}, 
\emph{Real Hilbert Subspaces, Modular Theory, SL$(2,\RR)$ and CFT}, 
\emph{\href{}{Ser.~Adv.~Math.~Sibiu}} {\bf 13} (2008) 33-91
Theta Foundation 

\bibitem{Lo21}

R. Longo,
\emph{Modular Structure of the Weyl Algebra},
\emph{\href{https://doi.org/10.1007/s00220-022-04344-7}{Commun.~Math.~Phys.}} {\bf 401}(1) (2023) 1061
[{\tt \href{https://arxiv.org/abs/2111.11266}{arXiv:2111.11266}}].

\bibitem{Maj88}

S. Majid, 
\emph{Hopf Algebras for Physics at the Planck Scale},  
\emph{\href{https://doi.org/10.1088/0264-9381/5/12/010}{Class.~Quantum Gravity}} {\bf 5} (1988) 1587-1606. 

\bibitem{Maj91}

S. Majid, 
\emph{Principle of Representation-theoretic Self-duality}, 
\emph{\href{https://doi.org/10.4006/1.3028923}{Phys.~Essays}} {\bf 4}(3) (1991) 395-405. 

\bibitem{Ma03}

D. Mauro, 
\emph{A New Quantization Map}, 
\emph{\href{https://doi.org/10.1016/s0375-9601(03)00996-4}{Phys.~Lett.~A}} {\bf 315}(1-2) (2003)28-35
\newline
[{\tt \href{https://arxiv.org/abs/quant-ph/0305063}{arXiv:quant-ph/0305063}}].

\bibitem{Na25}

J. Naudts, 
\emph{A Complex Structure for Two-Typed Tangent Spaces}, 
\emph{\href{https://doi.org/10.3390/e27020117}{Entropy}} {\bf 27}(2) (2025) 117.

\bibitem{NaZh25}

J. Naudts, J. Zhang, 
\emph{Pairs of Subspaces, Split Quaternions and the Modular Operator}, 
\emph{\href{https://doi.org/10.1016/s0034-4877(25)00058-8}{Rep.~Math.~Phys.}} {\bf 96}(1) (2025) 115-145
[{\tt \href{https://arxiv.org/abs/2501.04010}{arXiv:2501.04010}}].

\bibitem{Ne25}

K.-H. Neeb, 
\emph{Nets of Real Subspaces on Homogeneous spaces and Algebraic Quantum Field Theory},  
[{\tt \href{https://arxiv.org/abs/2511.09360}{arxiv:2511.09360}}] 2025.

\bibitem{Ne26}

K.-H. Neeb,
\emph{A Classification of Coadjoint Orbits Carrying Gibbs Ensembles},  
\newline
[{\tt \href{https://arxiv.org/abs/2601.04934}{arXiv:2601.04934}}] 2026.

\bibitem{Oj81}

I. Ojima,
\emph{Gauge Fields at Finite Temperatures-`Thermo Field Dynamics’ and the KMS Condition and Their Extension to Gauge Theories}, 
\emph{\href{https://doi.org/10.1016/0003-4916(81)90092-0}{Ann.~Phys.}} {\bf 137}(1) (1981) 1-32. 

\bibitem{Pu21}

C. Puttirunjroj,
\emph{Horizontal Categorification of Spectral Triples},
\emph{\href{10.14457/TU.the.2021.472}{Ph.D.~Thesis}} (2021) Thammasat University.  

\bibitem{Ra17a}

M. Raasakka, 
\emph{Spacetime-Free Approach to Quantum Theory and Effective Spacetime Structure},
\emph{\href{https://doi.org/10.3842/sigma.2017.006}{SIGMA}} {\bf 13} (2017) 006 
[{\tt \href{https://arxiv.org/abs/1605.03942}{arXiv:1605.03942}}].

\bibitem{Ra17b}

M. Raasakka,
\emph{Local Lorentz Covariance in Finite-dimensional Local Quantum Physics}, 
\emph{\href{https://doi.org/10.1103/physrevd.96.086023}{Phys.~Rev.~D}} {\bf 96} (2017) 086023 
[{\tt \href{https://arxiv.org/abs/1705.06711}{arXivL1705.06711}}].

\bibitem{RS61}

H. Reeh, S. Schlieder,  
\emph{Bemerkungen zur Unit\"ar\"aquivalenz von Lorentzinvarianten Feldern},  
\emph{\href{https://doi.org/10.1007/BF02787889}{Nuovo Cim.}} {\bf 22} (1961) 1051-1068. 

\bibitem{RvD77}

M. A. Rieffel, A. van Daele, 
\emph{A Bounded Operator Approach to Tomita-Takesaki Theory}, 
\emph{\href{https://doi.org/10.2140/pjm.1977.69.187}{Paciﬁc J.~Math.}} {\bf 69}(1) (1977) 187-221. 

\bibitem{Sa67} 

A. Sakharov, 
\emph{Violation of CP Invariance, C Asymmetry, and Baryon Asymmetry of the Universe}, 
\href{https://doi.org/10.1070/PU1991v034n05ABEH002497}{\emph{JETP Letters}} {\bf 5} (1967) 24-27.  

\bibitem{Sa80}

A. Sakharov, 
\emph{Cosmological Models of the Universe with Reversal of Time's Arrow}, 
\href{https://doi.org/10.3367/UFNr.0161.199105k.0081}{\emph{JETP}} {\bf 79} (1980) 3(9) 689-693. 

\bibitem{Str66}

R. F. Streater,
\emph{Canonical Quantisation}, 
\emph{\href{https://doi.org/10.1007/bf01773360}{Commun.~Math.~Phys.}} {\bf 2} (1966) 354-374.

\bibitem{Su06}

S. J. Summers, 
\emph{Tomita-Takesaki Modular Theory}, 
in \emph{\href{https://doi.org/10.1016/b0-12-512666-2/00019-5}{Encyclopedia of Mathematical Physics}}, 251-257
Elsevier 2006
[{\tt \href{https://arxiv.org/abs/math-ph/0511034}{arXiv:math-ph/0511034}}].

\bibitem{Ta03}

M. Takesaki,
\emph{Theory of Operator Algebras. II},  
Operator Algebras and Non-commutative Geometry 6,  
Springer 2003.

\bibitem{Tul74}

W. M. Tulczyjew, 
\emph{Hamiltonian Systems, Lagrangian Systems, and the Legendre Transformation},  
\textit{Sympos.~Math.} {\bf XIV} (1974) 247-258.  

\bibitem{VdB04}

M. Van den Bergh,
\emph{Double Poisson Algebras}, 
\emph{\href{https://doi.org/10.1090/S0002-9947-08-04518-2}{Trans.~Am.~Math.~Soc.}} {\bf 360}(11) (2008) 5711-5769
[{\tt \href{https://arxiv.org/abs/math/0410528}{arXiv:math/0410528}}].

\bibitem{VeZu19a}

E. P. Verlinde, K. M. Zurek, 
\emph{Observational Signatures of Quantum Gravity in Interferometers}, 
\emph{\href{https://doi.org/10.1016/j.physletb.2021.136663}{Phys.~Lett.~B}} {\bf 822} (2021) 136663
[{\tt \href{https://arxiv.org/abs/1902.08207}{arxiv:1902.08207}}].

\bibitem{VeZu19b}

E. P. Verlinde, K. M. Zurek, 
\emph{Spacetime Fluctuations in AdS/CFT}, 
\emph{\href{https://doi.org/10.1007/jhep04(2020)209}{J.~High Energy Phys.}} {\bf 04} (2020) 209
[{\tt \href{https://arxiv.org/abs/1911.02018}{arXiv1911.02018}}].

\bibitem{VeZu22}

E. P. Verlinde, K. M. Zurek, 
\emph{Modular Fluctuations from Shockwave Geometries},
\emph{\href{https://doi.org/10.1103/physrevd.106.106011}{Phys.~Rev.~D}} {\bf 106}(10) (2022) 106011
[{\tt \href{https://arxiv.org/abs/2208.01059}{arxiv:2208.01059}}].

\bibitem{Wa94}

R. M. Wald, 
\emph{Quantum Field Theory in Curved Spacetime and Black Hole Thermodynamics}, 
Chicago University Press 1994.

\bibitem{Wei97}

A. Weinstein, 
\emph{The Modular Automorphism Group of a Poisson Manifold},  
\emph{\href{https://doi.org/10.1016/s0393-0440(97)80011-3}{J.~Geom.~Phys.}} {\bf 23}(3) (1997) 379-394 

\bibitem{Wor72}

S. L. Woronowicz, 
\emph{On the Purification of Factor States},
\emph{\href{https://doi.org/10.1007/bf01645776}{Commun.~Math.~Phys.}} {\bf 28} (1972) 221-235. 

\end{thebibliography}
\end{document}